%% file: icse27.tex
\documentclass[conference]{IEEEtran}
\IEEEoverridecommandlockouts

\usepackage{listings}
\usepackage{xcolor}
\usepackage{tcolorbox}
\usepackage[normalem]{ulem}

\usepackage{enumitem}

\usepackage{caption}

\usepackage{booktabs}

\usepackage{balance}
\AtBeginDocument{%
  }

\lstdefinelanguage{Python}{
    morekeywords={class, def, return, try, except, raise, from},
    keywordstyle=\color{blue},
    stringstyle=\color{green},
    commentstyle=\color{gray},
    morecomment=[l]{\#},
}

\lstdefinelanguage{diff}{
    morecomment=[f][\color{green}]{+},
    morecomment=[f][\color{red}]{-},
    morecomment=[f][\color{blue}]{@@},
}

\lstdefinelanguage{errorlog}{
    morecomment=[f][\color{red}]{E},
    morecomment=[f][\color{magenta}]{?},
}

\lstdefinestyle{yaml}{
     basicstyle=\color{blue}\footnotesize,
     rulecolor=\color{black},
     string=[s]{'}{'},
     stringstyle=\color{blue},
     comment=[l]{:},
     commentstyle=\color{black},
     morecomment=[l]{-}
}

\usepackage{algorithm}
\usepackage{algorithmic}

\usepackage{listings}
\usepackage{xspace}

\usepackage[T1]{fontenc}

\usepackage[utf8]{inputenc}

\usepackage{microtype}

\usepackage{graphicx}

\usepackage{subcaption}

\usepackage{soul}
\usepackage{framed}
\usepackage{multirow}
\usepackage{siunitx}
\newcommand*\colourcheck[1]{%
	\expandafter\newcommand\csname #1check\endcsname{\textcolor{#1}{\ding{52}}}%
}
\colourcheck{blue}
\colourcheck{green}
\colourcheck{red}

\newcolumntype{L}[1]{>{\raggedright\arraybackslash}p{#1}}

\newcommand{\code}[1]{{\footnotesize\texttt{#1}}}
\usepackage{amsthm}
\definecolor{dkgreen}{rgb}{0,0.6,0}
\definecolor{gray}{rgb}{0.5,0.5,0.5}
\definecolor{lightgray}{rgb}{211, 211, 211}
\definecolor{mauve}{rgb}{0.58,0,0.82}

\newcommand{\tool}{\textsc{APIPilot}\xspace}

\newboolean{showcomments}
\setboolean{showcomments}{true}
\ifthenelse{\boolean{showcomments}}
 { \newcommand{\mynote}[2]{
      \fbox{\bfseries\sffamily\scriptsize#1}
        {\small$\blacktriangleright$\textsf{\emph{#2}}$\blacktriangleleft$}}}
        { \newcommand{\mynote}[2]{}}

\newcolumntype{L}[1]{>{\raggedright\arraybackslash}p{#1}}
\usepackage{amsthm}
\definecolor{dkgreen}{rgb}{0,0.6,0}
\definecolor{gray}{rgb}{0.5,0.5,0.5}
\definecolor{lightgray}{rgb}{211, 211, 211}
\definecolor{mauve}{rgb}{0.58,0,0.82}

\definecolor{custom-red}{rgb}{0,0,0}
\definecolor{custom-blue}{rgb}{0,0,0}

\definecolor{c1}{HTML}{f4cccc}
\definecolor{c2}{HTML}{f5cdcd}
\definecolor{c3}{HTML}{fffcfc}
\definecolor{c4}{HTML}{ffffff}
\definecolor{c5}{HTML}{ffffff}
\definecolor{c6}{HTML}{fffdfd}
\definecolor{c7}{HTML}{f5cfcf}
\definecolor{c8}{HTML}{fffbfb}
\definecolor{c9}{HTML}{ffffff}
\definecolor{c10}{HTML}{fffdfd}
\definecolor{c11}{HTML}{fefafa}
\definecolor{c12}{HTML}{fef7f7}
\definecolor{c13}{HTML}{ffffff}
\definecolor{c14}{HTML}{fffefe}
\definecolor{c15}{HTML}{ffffff}
\definecolor{c16}{HTML}{fefafa}
\definecolor{c17}{HTML}{fdf3f3}
\definecolor{c18}{HTML}{fffefe}
\definecolor{c19}{HTML}{fdf5f5}
\definecolor{c20}{HTML}{ffffff}

\begin{document}


\title{REST API Testing with Verified LLM-Inferred Dependencies and Response-Driven Refinement}

\author{
\IEEEauthorblockN{Tu Nguyen}
\IEEEauthorblockA{\textit{University of Science, VNU-HCM}\\
Ho Chi Minh City, Vietnam\\
23C15017@student.hcmus.edu.vn}
\and
\IEEEauthorblockN{Thanh Nguyen}
\IEEEauthorblockA{\textit{University of Science, VNU-HCM}\\
Ho Chi Minh City, Vietnam\\
22127389@student.hcmus.edu.vn}
\and
\IEEEauthorblockN{Huy Nguyen}
\IEEEauthorblockA{\textit{University of Science, VNU-HCM}\\
Ho Chi Minh City, Vietnam\\
22127154@student.hcmus.edu.vn}
\and
\IEEEauthorblockN{Viet Nguyen}
\IEEEauthorblockA{\textit{University of Science, VNU-HCM}\\
Ho Chi Minh City, Vietnam\\
21C11045@student.hcmus.edu.vn}
\and
\IEEEauthorblockN{Tien N. Nguyen}
\IEEEauthorblockA{\textit{University of Texas at Dallas}\\
Richardson, TX, USA\\
tien.n.nguyen@utdallas.edu}
\and
\IEEEauthorblockN{Vu Nguyen\textsuperscript{*}}
\IEEEauthorblockA{\textit{University of Science, VNU-HCM; Katalon LLC.}\\
Ho Chi Minh City, Vietnam\\
nvu@fit.hcmus.edu.vn}
\thanks{\textsuperscript{*}Corresponding author.}
}



\maketitle
\thispagestyle{plain} 

\begin{abstract}
Testing RESTful APIs requires generating sequences of API calls that
satisfy dependencies among operations, parameters, and runtime-created
resources. Recent LLM-based approaches infer such dependencies and
generate test sequences from OpenAPI specifications, but they often
treat LLM-inferred relationships as correct without execution-based
validation. This can introduce spurious dependencies, miss feasible
operation chains, and produce infeasible tests. In this paper, we
propose \textsc{APIPilot}, an execution-validated framework for REST
API testing. \textsc{APIPilot} first derives candidate producer-consumer
dependencies from OpenAPI specifications using structural heuristics
and LLM-based semantic reasoning. It then treats these dependencies as
hypotheses and validates them through concrete API executions before
using them for test generation. The validated dependencies are organized
into a dependency graph from which \textsc{APIPilot} constructs
coverage-aware workflows via bounded top-$k$ graph traversal, separating
semantic dependency inference from sequence construction. To improve
subsequent tests, \textsc{APIPilot} further performs response-driven
refinement: runtime responses are analyzed to update resource pools,
adjust input-generation constraints, and prune or revise invalid
dependency mappings. Empirical evaluation on 16 real-world REST API
services shows that \textsc{APIPilot} achieves 92.3\% operation
coverage, up to 58.6\% code coverage, and an 88.1\% workflow execution
success rate, outperforming both LLM-based and traditional REST API
testing baselines. \textsc{APIPilot} also detects 197 unique
5xx failures and specification--execution mismatches,
demonstrating the benefit of grounding dependency inference in
execution feedback.
\end{abstract}


\input{sections/introduction}
\input{sections/motivation}
\input{sections/key_idea}
\input{sections/new-arch}

\input{sections/evaluation}

\input{sections/threats_to_validity}

\input{sections/related}

\section{Conclusion}

We presented \tool, an execution-grounded framework for REST API testing. 
\tool\ treats inferred dependencies as hypotheses, validates them through concrete API executions, and constructs coverage-aware workflows from the validated dependency graph rather than relying on LLM-generated sequences. It further uses runtime responses to refine resource pools, input constraints, and dependency mappings across testing iterations. Our results show that grounding LLM-guided dependency inference in execution feedback improves workflow success, operation coverage, code coverage, and fault-revealing capability over LLM-based REST API testing approaches.

\section{Data Availability Statement}

Our data and code are publicly available at our website~\cite{api_pilot_replication_2026}.


\balance

\bibliographystyle{IEEEtran}

\bibliography{references}

\end{document}

%% file: sections/introduction.tex
\section{Introduction}
\label{sec:intro}

RESTful APIs become the dominant architectural style for web services \cite{fielding2000rest,neumann2018analysis}. Because serving as interfaces between independently developed components, their correctness, robustness, and security are critical to modern software reliability. Automated test generation for RESTful APIs has thus received substantial attention \cite{ehsan2022restful,golmohammadi2022testing}. Existing techniques include black-box and white-box approaches \cite{golmohammadi2022testing}. Black-box tools generate requests from API specifications such as OpenAPI, often using specification analysis, dynamic feedback, fuzzing, or search-based exploration~\cite{atlidakis2019restler,arcuri2018evomaster}. White-box techniques instead analyze server-side code, e.g., via symbolic execution or static analysis, to guide test generation \cite{godefroid2012sage}. Despite this progress, REST API testing remains difficult because many APIs expose stateful usage protocols: operations must be invoked in appropriate orders, and values produced by one operation are often required by later operations \cite{zhang2023open,kim2022automated}.

These dependencies arise at several levels. At the operation level, one endpoint may need to precede another, such as creating a resource before retrieving, updating, or deleting it. At the parameter level, inputs may need to satisfy constraints involving identifiers, tokens, foreign keys, enum values, or domain-specific formats \cite{martin2020automated}. Across operations, response fields produced by one operation may serve as required inputs to another. Accurately exercising such producer--consumer relationships is vital for constructing executable API workflows and reaching meaningful server-side behavior \cite{liu2022morest,kim2022automated}. However, existing techniques often rely on heuristics, partial dynamic feedback, or search-based exploration, which can miss implicit dependencies or retain spurious ones.

Recent work has explored large language models (LLMs) for REST API testing
\cite{kim2024leveraging,kim2025autoresttest,kogler2025restifai}. 
KAT \cite{le2024kat} uses LLMs to infer operation and inter-parameter
dependencies and to generate test sequences from OpenAPI specifications.
AutoRestTest~\cite{kim2025autoresttest} constructs a dependency graph from the
specification and uses LLM-guided agents to produce test sequences.
These techniques show that LLMs can help interpret natural-language endpoint
descriptions, parameter names, and schema information that are difficult to
capture using purely syntactic rules.

However, {\em directly trusting LLM-inferred dependencies can make REST API testing
fragile}. An LLM may infer a~depen\-dency between two operations because their
names or~desc\-riptions appear related, even when values produced by the first
operation cannot actually satisfy the second operation at runtime. Conversely,
it may miss feasible operation chains when the dependency is implicit in the
API behavior rather than explicitly documented in the specification. Such
errors can lead to hallucinated dependency edges, infeasible request chains,
and repeated 4xx responses that provide little coverage of meaningful
server-side behavior \cite{zhang2025hallucinations}. Generally, an LLM
prediction should be treated as a candidate hypothesis rather than as evidence
that a dependency is useful for testing.

This paper proposes \tool, an {\bf execution-validated} framework for REST API
testing. \tool\ first infers candidate producer--consumer dependencies from
OpenAPI specifications using structural heuristics and LLM-based semantic
reasoning. Rather than directly relying on these inferred dependencies, \tool\
validates them through concrete API executions and retains only dependencies
that can support executable request chains under the current testing context.
The validated dependencies are organized into a dependency graph, which is then
used to construct {\bf coverage-aware workflows} via bounded top-$k$ graph traversal.
This design separates semantic dependency inference from sequence construction:
LLMs help propose candidate dependencies, while workflow generation is performed
systematically over execution-validated evidence.

\tool\ further improves testing through {\bf response-driven refinement}. During
execution, \tool\ records status codes, response bodies, and extracted resource
values in a contextual resource pool. Runtime responses are then used to refine
subsequent testing by updating (in)valid value pools, adjusting
input-generation constraints, and pruning or revising dependency mappings that
lead to repeated failures. Deterministic checks derived from the schema
are used for syntactic constraints, e.g., required fields, types, formats,
enums, and documented status codes, while LLM reasoning is reserved for
under-specified semantic constraints and treated as refinement guidance rather
than as an oracle. A successful 2xx response is thus interpre\-ted as
evidence that a dependency is executable in current~context.


We evaluate \tool\ on 16 REST API services against four baseline tools: DeepREST~\cite{corradini2024DeepREST}, a state-of-the-art non-LLM approach, and three recent LLM-based tools: AutoRestTest, RESTifAI, and KAT. \tool\
achieves the highest average operation coverage (92.3\%), outperforming DeepREST (60.4\%),
AutoRestTest (70.4\%), RESTifAI (79.0\%), and KAT (68.1\%). It also obtains the
best average code coverage among the instrumented services (47.0\% branch
coverage, 58.6\% line coverage, and 58.6\% method coverage), the highest
workflow execution success rate (88.1\%), and a competitive token cost of 155
tokens per generated test case. \tool\ also detects 197 unique
5xx failures and specification--execution mismatches, demonstrating the
benefit of grounding LLM-guided dependency inference in execution feedback.

In summary, this paper makes the following contributions:

\begin{itemize}
    \item We propose an {\bf execution-validated dependency inference framework} for
    REST API testing that treats LLM-inferred producer--consumer relationships
    as hypotheses and validates them through concrete API executions.

    \item We design a {\bf coverage-aware workflow construction algorithm} that
    derives bounded top-$k$ executable workflows from validated dependencies,
    separating semantic dependency inference from test sequence generation.

    \item We introduce {\bf response-driven refinement} that uses runtime feedback to
    update resource pools, input-generation constraints, and revise invalid dependency~mappings.

    \item We conduct extensive empirical evaluation on {\tool}.
\end{itemize}





%% file: sections/motivation.tex
\section{Motivation}
\label{sec:motiv}

\subsection{Motivating Example}

\begin{figure}[t]
\centering
\lstset{
		numbers=left,
		numberstyle= \tiny,
		keywordstyle= \color{blue!70},
		commentstyle= \color{red!50!green!50!blue!50},
		frame=shadowbox,
		rulesepcolor= \color{red!20!green!20!blue!20} ,
		xleftmargin=1.5em,xrightmargin=0em, aboveskip=1em,
		framexleftmargin=1.5em,
        numbersep= 5pt,
		language=Java,
        basicstyle=\scriptsize\ttfamily,
        numberstyle=\scriptsize\ttfamily,
        emphstyle=\bfseries,
        moredelim=**[is][\color{red}]{@}{@},
		escapeinside= {(*@}{@*)}
	}
\begin{lstlisting}[style=yaml] 
openapi: 3.0.1
info:
    title: GitLab Commit API
paths:
  "/projects":
    get:
      summary: Get a list of all visible projects...
    post:
      summary: Create a project owned by the authenticated user.
  "/projects/{id}/repository/commits":
    get:
      summary: Get a list of repository commits in a project.
    post:
      summary: Create a commit by posting a JSON payload.
  "/projects/{id}/repository/commits/{sha}":
    get:
      summary: Get a specific commit identified...
  ...
components: ...
\end{lstlisting}
\vspace{-9pt}
\caption{Endpoints of GitLab's Repository Commits API}
\label{fig:gitlab}
\vspace{-12pt}
\end{figure}

We use GitLab's Repository Commits API~\cite{gitlab} to illustrate the
challenges of dependency-aware REST API testing. In
Fig.~\ref{fig:gitlab}, we have API operations for listing~projects,
commits in a project, creating commits, and retrieving a
commit. To test the endpoint
\code{GET /projects/\{id\}/repository/commits/\{sha\}}, a tool must construct
an executable workflow that supplies both the project \code{id} and
commit identifier \code{sha}. A feasible {\bf workflow}:

\code{GET /projects} $\rightarrow$

\code{GET /projects/\{id\}/repository/commits} $\rightarrow$

\code{GET /projects/\{id\}/repository/commits/\{sha\}}.

In this workflow, \code{GET /projects} produces a project~\code{id}, which can
be used as the path parameter of
\code{GET /projects/ \{id\}/repository/commits}. The latter operation returns
commit objects containing a \code{sha}, which can then be used to retrieve a
specific commit. This shows that useful REST API tests require
producer-consumer reasoning across multiple operations.

However, dependency inference from an OpenAPI specification is error-prone.
\underline{First}, an approach may infer unnecessary ordering constraints. For example,
RESTifAI~\cite{kogler2025restifai} and KAT~\cite{le2024kat} may require
\code{POST /projects} before testing \code{GET /projects}, although
\code{GET /projects} can be executed independently and can already provide
valid project identifiers. Such a dependency unnecessarily lengthens the
workflow and may make later operations unreachable if project creation fails.

\underline{Second}, similarity-based dependency inference may retain spurious edges. For
example, AutoRestTest~\cite{kim2025autoresttest} constructs a Semantic Property
Dependency Graph using embedding similarity. This may incorrectly map
\code{short\_id}~from
\code{GET /projects/\{id\}/repository/commits} to the unrelated
\code{id\_after} parameter of \code{GET /projects}, while missing the more
useful dependency from commit \code{sha} to
\code{GET /projects/ \{id\}/repository/commits/\{sha\}}. Such edges can produce
syntactically valid but semantically ineffective request chains.

\underline{Third}, failed requests require more precise feedback than a status code alone.
For example, a 404 response may be caused by an invalid project \code{id}, a
missing commit \code{sha}, or a faulty dependency mapping that injects a value
from the wrong source. Treating all 404 responses as generic invalid-input
failures can lead a tool to repeatedly mutate values without correcting the
underlying dependency. Similarly, required fields and schema formats can often
be checked deterministically from the OpenAPI specification, whereas
under-specified semantic constraints require additional runtime evidence.

These observations motivate a design principle: LLMs should help propose
candidate dependencies, but their outputs should not be trusted as final test
logic. Instead, candidate dependencies should be treated as hypotheses,
{\bf validated through concrete API execution, and refined using runtime feedback}.

%% file: sections/key_idea.tex
\subsection{Key Ideas}
\label{sec:ideas}

We design \tool\ around the following key ideas.

\vspace{1pt}
\textbf{\textit{Key Idea 1: Validate LLM-inferred dependencies by~concrete execution.}}
Instead of directly accepting dependencies inferred from names, descriptions,
or embeddings, \tool\ represents each dependency as an executable candidate
producer-consumer mapping. In our example, the~candidate edge from
\code{GET /projects} to
\code{GET /projects/\{id\}/repository /commits} records that the response field
\code{id} may supply the path parameter \code{id}. 
Each edge thus carries not only an operation pair, but also the specific source
field, target parameter, dependency type, and evidence used to infer it.
A candidate dependency is retained only when it can support an executable
request chain. 


\vspace{2pt}
\textbf{\textit{Key Idea 2: Generate coverage-aware workflows by graph
traversal, not by LLM sequence synthesis.}}
After dependency validation, \tool\ organizes verified dependencies into an
Operation Dependency Graph (ODG) and derives executable workflows through bounded
top-$k$ graph traversal. 

%

It can also avoid unnecessary workflows that start with \code{POST /projects}
when \code{GET /projects} already provides valid project identifiers. The workflow construction rather than API execution is deterministic. Given the same validated graph, ranking function, depth bound,
top-$k$ value, random seed, and initial service state, \tool\ produces the same
workflow set. However, the actual API responses may still vary if the
black-box service state changes outside \tool's control.

\vspace{2pt}
\textbf{\textit{Key Idea 3: Refine values and dependencies from runtime
responses.}}
During execution, \tool\ records response values in a contextual resource pool
and updates this pool using runtime feedback. Successful responses add
validated resources, such as project \code{id}s and commit \code{sha}s, to a
whitelist for later reuse. Repeated failures add invalid values or mappings to
a blacklist. For example, if a workflow repeatedly produces 404 responses when
a value extracted from one field is injected into an unrelated parameter,
\tool\ can suppress that value source or revise the corresponding dependency
edge. 


\tool\ refines test generation through a layered feedback process, instead of
delegating all response analysis to LLM. \underline{First}, {\em deterministic OpenAPI
validation} checks schema-level properties, e.g., required fields, primitive
types, formats, enums, ranges, documented status codes, and response schemas.
\underline{Second}, {\em execution evidence} from status codes, response bodies, extracted
resources, and repeated successes or failures is used to update whitelists,
blacklists, parameter constraints, and dependency-edge statuses. \underline{Third}, {\em LLM} is invoked only for under-specified semantic constraints or ambiguous
root-cause hypotheses that cannot be directly resolved from the specification.
Its output is treated as refinement guidance, not as an oracle, and is applied
only when validated by later~execution.



%% file: sections/new-arch.tex
\section{{\tool} Workflow}

\begin{figure}[t]
  \centering
  \includegraphics[width=4in]{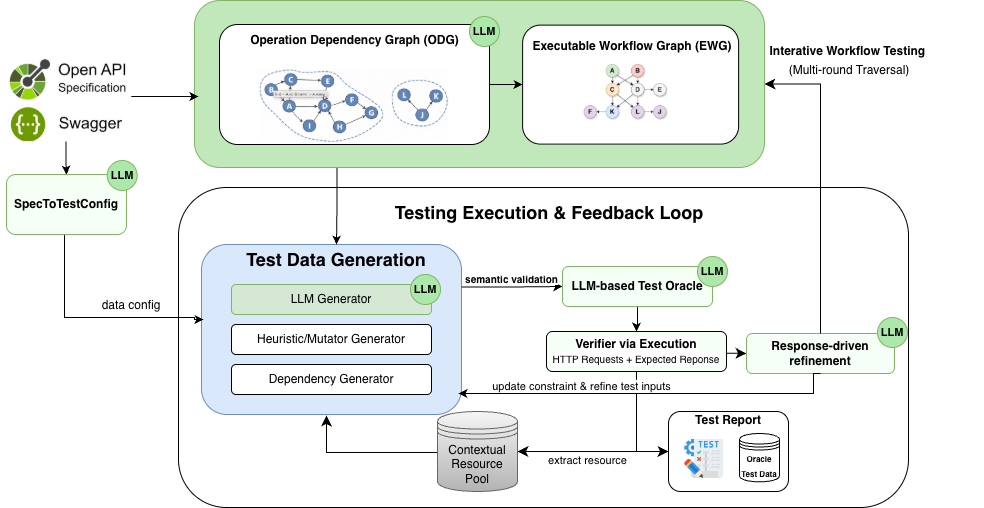} 
  \vspace{-12pt}
  \caption{APIPilot Architecture}
  \label{fig:architecture}
\end{figure}

Fig.~\ref{fig:architecture} displays {\tool}'s workflow.
The design~prinple is to validate inferred dependencies:
it first proposes producer-consumer dependencies from the 
specification, validates them through concrete executions, constructs
coverage-aware workflows from the validated dependencies, and refines input generation and dependency mappings via runtime responses.

We use the GitLab Commit API in Fig.~\ref{fig:gitlab} as a running example.
To test
\code{GET /projects/\{id\}/repository/commits/ \{sha\}},
a tool must obtain a valid project \code{id} and a commit \code{sha}. One
feasible workflow is as the one in the motivating example.




Here, \code{GET /projects} produces a project \code{id}; the commit-listing
operation consumes this \code{id} and produces commit objects containing
\code{sha}; the final operation consumes both values to retrieve a specific
commit. \tool\ is designed to discover, validate, and exploit such workflows
without blindly trusting either heuristic matches or LLM-inferred dependencies.

\subsection{Overview}

\begin{algorithm}[t]
\footnotesize
\caption{\tool\ Overview}
\label{alg:apipilot_overview}
\begin{algorithmic}[1]
\STATE \textbf{Input:} Specification $S$, testing budget $B$, depth limit $d_{\max}$, top-$k$
\STATE \textbf{Output:} Executed tests, validated dependencies, detected failures
\STATE Parse $S$ to obtain operations, parameters, schemas, and responses
\STATE Infer candidate dependencies with structural heuristics and LLM reasoning
\STATE Initialize the Contextual Resource Pool and per-field test configurations
\FOR{each testing iteration within budget $B$}
    \STATE Validate candidate dependencies through concrete executions
    \STATE Update the ODG with validated, rejected, and refined dependencies
    \STATE Construct top-$k$ workflows from the validated ODG using bounded traversal
    \FOR{each workflow}
        \STATE Generate request inputs using schema rules, mutation, and resource-pool values
        \STATE Execute the request sequence, record status codes and responses
        \STATE Extract reusable resource values into the Contextual Resource Pool
        \STATE Refine value pools, parameter constraints, and dependency mappings from responses
    \ENDFOR
\ENDFOR
\STATE \textbf{return} executed tests, refined ODG, and detected failures
\end{algorithmic}
\end{algorithm}

Algorithm~\ref{alg:apipilot_overview} summarizes the end-to-end procedure of
\tool. Details on each component are given later.

\subsubsection{\bf Candidate dependency inference}
\tool\ parses the OpenAPI specification to extract operations, path/query/body
parameters, request schemas, response schemas, status codes, and textual
descriptions. It then constructs candidate producer--consumer dependencies
using two complementary sources of evidence. Structural heuristics capture
common REST patterns, such as identifier propagation, shared path templates,
resource hierarchies, lifecycle patterns, and hypermedia links. LLM-based
semantic reasoning is used to analyze cases that are difficult to resolve from
syntax alone, such as when parameter names and response fields are semantically
related but not lexically identical.
For example, in the GitLab API, structural analysis can identify that the
response field \code{id} from \code{GET /projects} is a plausible source for
the path parameter \code{id} in
\code{GET /projects/\{id\}/repository/commits}. Similarly, semantic reasoning
can help identify that the \code{sha} field returned by the commit-listing
operation is the appropriate value for the \code{sha} parameter in
\code{GET /projects/\{id\}/repository/commits/\{sha\}}.

\subsubsection{\bf Execution validation}
Candidate dependencies are not immediately used as final test logic. Instead,
\tool\ validates them by executing concrete API calls and checking whether the
produced values can support the consumer operation. In our example,
\tool\ executes \code{GET /projects}, extracts a concrete project \code{id},
injects it into
\code{GET /projects/\{id\}/repository/commits}, and observes whether the
operation can execute under the generated context. If successful, the
dependency from project \code{id} to the path parameter \code{id} is promoted
from a candidate dependency to a validated dependency. In contrast, a spurious
mapping such as using commit \code{short\_id} as the \code{id\_after}
parameter of \code{GET /projects} is not promoted unless execution
supports it.

\subsubsection{\bf Coverage-aware workflow construction} After validation, \tool\ organizes the validated dependencies into an ODG. Each node is an API operation, and each edge records a
validated producer--consumer relationship between a response field and a target
parameter. \tool\ then constructs executable workflows by traversing this graph
backward from each target operation and retaining the top-$k$ shortest and most
feasible workflows under a depth bound. This separates dependency inference
from sequence generation: LLMs help propose candidate dependencies, but
workflows are generated systematically from validated graph evidence.
In our example, the workflow targeting
\code{GET /projects/\{id\}/}\code{repository/commits/\{sha\}} is derived by resolving
its required inputs. The \code{sha} parameter can be supplied by
\code{GET /projects/\{id\}/repository/commits}, whose \code{id} parameter can
in turn be supplied by \code{GET /projects}. Thus, \tool\ constructs the
workflow as in the motivation section.
If \code{GET /projects} already provides valid project IDs, it
does not need to force an unnecessary prefix such as \code{POST /projects}.

\subsubsection{\bf Response-driven refinement}
During execution, \tool\ records responses and uses them to refine both value
generation and dependency mappings. Successful responses contribute reusable
values, such as project \code{id}s and commit \code{sha}s, to a contextual
resource pool. Repeated failures are used to blacklist invalid values,
deprioritize unreliable sources, or prune faulty dependency edges. For example,
a 404 response from
\code{GET /projects/\{id\}/repository/commits/\{sha\}} may indicate that the
project \code{id} is invalid, the commit \code{sha} does not belong to that
project, or the dependency mapping injected a value from the wrong response
field. \tool\ distinguishes these cases using the request, response,
specification constraints, and dependency provenance, and updates subsequent
tests.


\subsection{Dependency Representation}

The ODG is not intended to introduce a new graph abstraction; rather, it is the
internal representation used by \tool\ to store dependency evidence. Let
$\mathcal{O}$ be the set of operations extracted from the OpenAPI
specification. Each operation $o \in \mathcal{O}$ has input parameters
$P_o$ and response fields $R_o$. A dependency edge is represented as:
\[
e = \langle o_s, o_t, r, p, type, evidence, status \rangle ,
\]
where $o_s$ is the source operation, $o_t$ is the target operation,
$r \in R_{o_s}$ is the producer response field, $p \in P_{o_t}$ is the
consumer parameter, \textit{type} records the dependency category,
\textit{evidence} records how the dependency was inferred, and
\textit{status} records whether the edge is candidate, validated, rejected, or
refined.

This representation is more precise than a plain operation-level edge. For
example, the edge

\noindent 
$\langle$
\code{GET /projects},
\code{GET /projects/\{id\}/repository/commits},
\code{id},
\code{id},
\textit{identifier},
\textit{heuristic+execution},
\textit{validated}
$\rangle$

states that the response field \code{id} produced by \code{GET /projects}
supplies the path parameter \code{id} of the commit-listing operation. It does
not claim that every value produced by \code{GET /projects} can be used by the
target operation.

A candidate dependency is created only when the producer field and consumer
parameter are compatible with evidence:
$$
cand(e) \iff compatible(r,p) \wedge evidence(o_s,o_t,r,p).
$$
Here, $compatible(r,p)$ checks schema compatibility (type,
format, enum, path context, and identifier structure). The~predicate
$evidence(o_s,o_t,r,p)$ captures structural, semantic, lifecycle, hypermedia,
or LLM-based evidence showing that $r$ supplies~$p$.

A candidate dependency becomes validated only after execution provides evidence
that a value produced by $o_s$ can support the execution of $o_t$, that is, $valid(e) \iff $
\[
\exists v \in Val(o_s,r):
Exec(o_t[p \leftarrow v]) \in Success \cup Progress .
\]
$Val(o_s,r)$ denotes values extracted from field $r$ in responses of $o_s$.
$Success$ usually corresponds to documented 2xx responses. $Progress$ captures
cases where the injected value moves the request beyond a missing-resource or
missing-parameter failure, even if another independent constraint still causes
the request to fail. Thus, validation provides evidence that a dependency is
executable in the current testing context; it is not a proof of full
business-level correctness.

We further distinguish validated dependencies as "required" or "optional". 
A validated dependency is \emph{required} if the target
parameter $p$ is required by the OpenAPI specification or if repeated
executions show that omitting or replacing $p$ prevents progress. It is
\emph{optional} if the produced value can improve realism or coverage but the
target operation can still execute through another value source or without that
parameter. Lifecycle evidence is never treated as required by itself; it only
prioritizes candidate orderings until a concrete producer-consumer mapping is
validated by execution.

We consider {\em four dependency types}. An \textit{identifier dependency}
connects a created resource ID to a required input ID. A
\textit{schema dependency} connects a response attribute to a compatible input
field. A \textit{lifecycle dependency} captures REST ordering evidence, e.g.,
\code{POST} before \code{GET}, \code{PUT}, or \code{DELETE} on the same
resource; it is treated as candidate evidence rather than sufficient proof of a
valid dependency. A \textit{hypermedia dependency} is derived from explicit
links in responses when available.

\subsection{Candidate Dependency Inference}

\tool\ infers candidate dependencies using structural and semantic evidence.
The structural phase applies deterministic rules over paths, methods,
parameters, and schemas. It matches response fields to target parameters using
exact and normalized names, such as \code{id}, \code{project\_id}, and
\code{projectId}; checks type compatibility between producer fields and
consumer parameters; identifies hierarchical path relationships, such as
\code{/projects} and \code{/projects/\{id\}/repository/commits}; and records
REST lifecycle patterns and hypermedia links when present. To~avoid excessive
dependencies, generic fields such as \code{name}, \code{text},
\code{description}, etc. are not retained unless supported by stronger evidence, e.g., type compatibility, schema, or execution.

The semantic phase targets dependencies that structural rules may miss. \tool\
first uses an embedding model to retrieve semantically similar parameter--field
pairs from the specification. The LLM then examines the operation descriptions,
parameter descriptions, schema context, and candidate pair to decide whether the
pair plausibly represents a producer--consumer relationship. The LLM output is
constrained to a structured format that identifies the source operation, target
operation, source field, target parameter, dependency type, and explanation {\color{custom-blue}(the prompt templates provided in the replication package \cite{api_pilot_replication_2026})}.
Importantly, this output only creates a candidate edge; it does not by itself
validate the dependency.


\subsection{Execution Validation and Resource Pool}

The Contextual Resource Pool stores values extracted from successful API
responses. Each stored value is associated with its source operation, response
field, schema context, observed status code, and dependency provenance. During
execution, dependency-based generators retrieve values from this pool and
inject them into parameters of downstream operations.

To validate a candidate edge
$\langle$$o_s$, $o_t$, $r$, $p$, $type$, $evidence$, $candidate$$\rangle$, \tool\ executes
the source operation $o_s$,~extracts concrete values from field $r$, and uses
them to instantiate parameter $p$ of the target operation $o_t$. If the
operation succeeds under schema-valid inputs, or if the injected value enables
progress beyond a previous missing-resource/parameter failure, the
edge is marked as validated. If the same mapping repeatedly causes failures,
the edge is rejected or deprioritized.

A successful 2xx response is treated as evidence that the dependency is
executable in the current testing context, not as proof of full business-level
correctness. Conversely, a 4xx response does not automatically invalidate an
edge: it may result from an unrelated missing parameter, an invalid optional
field, authorization, or an inconsistent resource state. Therefore, \tool\
records the response together with the request and dependency provenance for
later refinement. 

\subsection{Executable Workflow Graph (EWG) Construction}

Once candidate dependencies are validated, \tool\ constructs workflows from the
validated ODG. For each target operation, it performs a bounded backward
traversal to resolve required inputs. Candidate predecessor operations are
ranked by dependency status, dependency type, source reliability, path
specificity, and workflow length. The traversal avoids cycles and retains only
the top-$k$ workflows per target operation.
We define the ranking of a workflow $W$ lexicographically as:

{\em 
rank(W)=$\langle$ unresolved(W), |W|, -validated(W), failed(W) $\rangle$.
}

Here, $unresolved(W)$ is the number of required inputs that remain unresolved,
$|W|$ is the workflow length, $validated(W)$ is the number of validated
dependency edges used in the workflow, and $failed(W)$ is the number of edges or
values previously associated with failures. \tool\ prefers workflows with fewer
unresolved inputs, shorter length, more validated dependencies, and fewer
failure-prone mappings. This ranking avoids introducing additional weighting
parameters and makes the top-$k$ selection deterministic.

Here, ``deterministic'' refers to workflow construction rather than API
execution. Given the same validated graph, ranking function, depth bound,
top-$k$ value, random seed, and initial service state, \tool\ produces the same
workflow set. However, the actual API responses may still vary if the
black-box service state changes outside its control. If several validated sources are available, \tool\ ranks them and retains the top-$k$ shortest workflows that cover the target operation.


\subsection{Test Data Generation}

For each operation in a workflow, \tool\ generates request inputs from a
per-field test configuration. This configuration is derived from the OpenAPI
specification and records the generation strategy for each parameter or request
body field. Schema-level constraints, such as required fields, primitive types,
formats, enums, array bounds, and numeric ranges, are handled deterministically.
For fields with well-defined constraints, \tool\ uses rule-based, boundary, or
mutation-based generators. For fields that depend on prior responses, it uses
the Dependency Generator to retrieve values from the Contextual Resource Pool.
For under-specified fields that cannot be instantiated from schema rules or
resource values, \tool\ may use an LLM value generator, but these generated
values are treated as candidates and refined through execution feedback.

The generator does not enumerate all possible values, which would be infinite.
Instead, it samples a finite set of field-presence combinations and value
strategies under a testing budget. For example, it can generate required-only
requests, required-plus-one-optional requests, all-field requests, boundary
cases, and intentionally invalid variants. A request is expected to succeed
when required fields are present, schema-level constraints are satisfied, and
dependency-sourced values are drawn from validated resource pools. A request is
expected to fail when \tool\ intentionally removes required fields, violates
schema constraints, mutates values outside allowed domains, or uses blacklisted
resource values. 

In the GitLab workflow, \tool\ uses the resource pool to fill the path
parameter \code{id} with a validated project identifier and \code{sha} with a
validated commit hash. Optional query parameters are then varied according to
their schema constraints and mutation budget. This allows \tool\ to exercise
both nominal scenarios and error-handling behavior without relying on an LLM to
synthesize entire requests from scratch.

\subsection{Response-driven Refinement}

After each request, we analyze the response and updates its testing state.
From Key Idea 3, response-driven refinement follows a three-layer design. 
{\bf First}, deterministic OpenAPI validation checks schema-level properties, including required fields, primitive types, formats, enums, ranges, documented status codes, and response schemas. {\bf Second}, execution evidence from status codes, response bodies, extracted resources, and repeated successes or failures is used to update resource whitelists, blacklists, parameter constraints, and dependency-edge statuses. {\bf Third}, LLM reasoning is invoked only for under-specified semantic constraints or ambiguous root-cause hypotheses that are not directly recoverable from the schema. LLM output is treated only as a candidate explanation.


\begin{itemize}
    \item \textbf{Schema violation}: the request violates a documented type,
    format, enum, range, or required-field constraint.

    \item \textbf{Invalid resource value}: a dependency-sourced value, e.g., a
    project \code{id} or commit \code{sha}, is rejected by the operation.

    \item \textbf{Faulty dependency mapping}: the source field is incompatible
    with the target parameter, even if their names or descriptions appear
    similar.

    \item \textbf{Missing or inconsistent context}: the target operation
    requires a resource state not established by the workflow.

    \item \textbf{Specification--execution mismatch}: the observed status code
    or response is inconsistent with the specification.
\end{itemize}

The resulting feedback is used to update both data generation and the ODG.
Successful values are added to whitelists and prioritized in later requests.
Values repeatedly associated with failures are blacklisted or deprioritized.
Parameter generators are tightened or relaxed based on observed constraints.
Candidate edges that repeatedly produce invalid requests are pruned, while new
producer--consumer relationships discovered from response bodies can be added
as candidates for validation.

In the GitLab example, suppose
\code{GET /projects/\{id\}/ repository/commits/\{sha\}} returns 404. \tool\
does not immediately treat the request as a generic invalid input. If the
project \code{id} was previously validated but \code{sha} came from a
different project, the refinement module attributes the failure to an
inconsistent resource pairing and records that \code{sha} values must be reused
with the project \code{id} from which they were obtained. If the \code{sha}
came from an unrelated response field, the corresponding dependency edge is
deprioritized or rejected. If the response reveals a new reusable identifier or
constraint, the resource pool and test configuration are updated for later.

%% file: sections/evaluation.tex
\section{Empirical Evaluation}
\label{sec:eval}

We conduct experiments to answer the following questions:

\begin{enumerate} [leftmargin=0.7em,labelsep=0.2em]

    \item \textbf{RQ1 (Effectiveness and Efficiency).} How effective and efficient is \tool\ in comparison with the baselines?

    \item \textbf{RQ2 (Workflow Exploration).} How effectively does {\tool} generate and execute API request sequences to explore API workflows compared to the baselines?

    \item \textbf{RQ3 (Error Detection).} How effective is {\tool} in generating tests to detect API faults?

    \item \textbf{RQ4 (Token and Cost Efficiency).} How does {\tool} achieve a balance between token efficiency and effectiveness?
    

\end{enumerate}
\subsubsection*{Experiment Setup}
{\color{custom-blue}\tool runs in fully automated mode without any user-supplied configuration overrides, ensuring a fair comparison with the fully automated baselines}.
All experiments are conducted on two machines, each equipped with an AMD Ryzen 7 3700X processor (3.6\,GHz, 8 cores, 16 logical processors), 16\,GB of RAM, and a GPU. To ensure consistent testing conditions, services are restarted and their databases restored at the beginning of each session, preventing state carryover across runs. All services run under default configurations and database settings. To avoid resource interference, each service and testing tool is allocated dedicated resources and executed sequentially. 


\subsubsection*{Baselines}

We compare \tool\ against 4 state-of-the-art REST API testing tools: {\bf AutoRestTest}~\cite{kim2025autoresttest}, {\bf RESTifAI}~\cite{kogler2025restifai}, {\bf KAT}~\cite{le2024kat}, and {\color{custom-red}DeepREST~\cite{corradini2024DeepREST}}. While {\color{custom-red}DeepREST} is a state-of-the-art non-LLM approach, the others are LLM-based approaches to REST API testing that collectively cover the primary paradigms: KAT uses GPT-based prompting to construct dependency graphs and generate test scripts; AutoRestTest combines LLMs with a semantic property dependency graph and multi-agent reinforcement learning; and RESTifAI employs LLMs to generate reusable, CI/CD-ready test workflows with error-driven retry. Moreover, their datasets are publicly available to enable a rigorous and reproducible comparison. To ensure a fair comparison, all tools are configured to use the same underlying LLM, OpenAI GPT-4.1-mini, with temperature set to 0 to minimize hallucinations and ensure deterministic output. Each experiment is repeated {\em five times}, and we report the average results to mitigate potential randomness.

\input{sections/datasets}

\input{sections/rq1_effectiveness}

\input{sections/rq2_workflow_exploration}
\input{sections/rq3_error_detection}
\input{sections/rq4_cost_efficiency}


%% file: sections/datasets.tex
\subsubsection*{Datasets}
\label{sec:datasets}

We evaluate \tool\ on a diverse benchmark of 16 real-world RESTful API services. The core dataset is drawn from established benchmarks used in prior work, including AutoRestTest, KAT and RESTifAI. 
We use these services because they already subsume the baselines' benchmarks, allowing direct and fair comparison  under identical experimental conditions used in these studies. To assess structural code coverage, we augment the benchmark with two open-source applications, \code{jhipster-app}~\cite{jhipster} and \code{petclinic-rest}~\cite{spring-petclinic}, 
for white-box coverage via JaCoCo~\cite{jacoco}, which
was reported only for the locally instrumentable open-source services.




\subsubsection*{Evaluation Metrics} We use metrics in three dimensions: 


\textbf{Operation Coverage (OC)} measures the percentage of API operations successfully invoked, i.e., those yielding at least one \texttt{2xx} response, relative to the total number of operations defined in the specification. It serves as the primary black-box indicator of a tool's ability to construct valid request sequences.

To evaluate response diversity beyond successful invocations, we measure \textbf{2xx Coverage} and \textbf{4xx Coverage}. Following the evaluation criteria in KAT~\cite{le2024kat}, 2xx Coverage captures the proportion of unique documented \texttt{2xx} status codes triggered by the test suite, reflecting the breadth of happy-path scenario coverage. 4xx Coverage measures the proportion of documented \texttt{4xx} codes triggered, assessing the ability to exercise client-side error handling, e.g., invalid inputs and constraint violations.

\input{tables/operation_coverages}

For white-box evaluation on our open-source services, we collect \textbf{Line Coverage (LC)}, \textbf{Branch Coverage (BC)}, and \textbf{Method Coverage (MC)} using JaCoCo, with all coverage metrics expressed as percentages. 
To measure test generation efficiency, we define the \textbf{Test Efficiency} ($\mathit{Eff}$) metric as:
\[
\mathit{Eff} = \frac{\#TC}{(OC/100 \times \#Ops)}
\]
where $\#TC$ is the total number of generated test cases, $OC$ is the achieved operation coverage percentage, and $\#Ops$ is the total number of operations in a service. A lower $\mathit{Eff}$ value indicates a more compact and less redundant test suite. 


To assess \textbf{fault detection} capability, we monitor \code{5xx} error responses, specification–execution mismatches, and undocumented status codes. Unlike \code{2xx} and \code{4xx} codes, which reflect expected API behaviors, \code{5xx} errors indicate unhandled server-side crashes or critical logical flaws.
normalized failure key. This key combines the API operation, HTTP method, status code, normalized error response, and response structure. Normalization removes volatile values such as identifiers, timestamps, random strings, etc.
Such mismatches capture discrepancies between the status codes listed in the specification and those observed at runtime, reflecting either execution failures or outdated specifications. Undocumented status codes are those from execution but not in the specification. 


%% file: tables/operation_coverages.tex
\begin{table*}[t]
\centering
\small
\setlength{\tabcolsep}{1pt}
\caption{{\color{custom-blue}Operation coverage (2xx) and efficiency comparison. 2xx/4xx: status code coverage(\%); Eff: test generation efficiency. {\bf GN}: Genome Nexus, {\bf LT}: Languagetool, {\bf RC}: Restcountries, {\bf JA}: jhipster-app, {\bf PR}: petclinic-rest.(RQ1)}}
\label{tab:operation_coverages}
\vspace{-6pt}
\begin{tabular}{lrr|rccc|rccc|rccc|rccc|rccc}
\toprule
\textbf{Service} & \textbf{\#Ops} & \textbf{LOC}
& \multicolumn{4}{c|}{\textbf{{\color{custom-red}DeepREST}}}
& \multicolumn{4}{c|}{\textbf{AutoRestTest}}
& \multicolumn{4}{c|}{\textbf{RESTifAI}}
& \multicolumn{4}{c|}{\textbf{KAT}}
& \multicolumn{4}{c}{\textbf{\tool}} \\
& & & \#TC & 2xx & 4xx & Eff.
      & \#TC & 2xx & 4xx & Eff.
      & \#TC & 2xx & 4xx & Eff.
      & \#TC & 2xx & 4xx & Eff.
      & \#TC & 2xx & 4xx & Eff. \\
\midrule
GN   & 23  & 20k
  & 2,554 & 82.6 & 4.3 & 111.0 & 4,000   & \textbf{100} & \textbf{100} & 173.9
  & 211     & 64.3         & 12.6         & \textbf{14.3}
  & 287     & 87.0         & 62.6         & \textbf{14.3}
  & 2,374   & \textbf{100} & \textbf{100} & 103.2 \\

LT   & 2   & 241k
  & 200 & 0.0 & 0.0 & 100.0 & 100,000 & \textbf{100} & \textbf{100} & 50000.0
  & 33      & \textbf{100} & \textbf{100} & 16.5
  & 9       & \textbf{100} & \textbf{100} & \textbf{4.5}
  & 218     & \textbf{100} & \textbf{100} & 109.0 \\

RC  & 22  & 2.4k
  & 3,912 & 86.4 & 90.9 & 177.8 & 150,000 & \textbf{100} & \textbf{100} & 6818.2
  & 211     & 91.0         & 92.7         & 10.5
  & 134     & 80.8         & 86.9         & \textbf{7.5}
  & 3,745   & \textbf{100} & \textbf{100} & 170.2 \\

JA   & 38  & 6.3k
  & 18,070 & 2.6 & 21.1 & 475.5 & 200,000 & \textbf{90.0} & \textbf{100} & 5848.0
  & 379     & 80.0          & \textbf{100} & \textbf{12.5}
  & 530     & 61.6          & 86.8         & 22.7
  & 5,550   & 89.5          & \textbf{100} & 163.3 \\

PR & 35  & 5.9k
  & 12,570 & 34.3 & 41.9 & 359.1 & 200,000 & 69.7          & \textbf{66.8} & 8205.5
  & 401     & \textbf{93.7} & 61.5          & \textbf{12.2}
  & 387     & 66.3          & 48.5          & 16.7
  & 6,864   & 82.9          & 54.3          & 236.7 \\

\midrule
Canada.H.    & 4   & --
  & 800 & \textbf{100} & 75.0 & 200.0 & 13,000  & \textbf{100} & \textbf{100} & 3250.0
  & 46      & \textbf{100} & \textbf{100} & \textbf{11.5}
  & 51      & \textbf{100} & \textbf{100} & 12.8
  & 350     & \textbf{100} & \textbf{100} & 87.5 \\

Bills API      & 21  & --
  & 264 & 38.1 & 41.2 & 12.6 & 8,000   & 77.2         & \textbf{80.9} & 493.8
  & 200     & 76.0         & 57.4          & \textbf{12.5}
  & 199     & 76.2         & 78.7          & \textbf{12.5}
  & 2,620   & \textbf{85.7} & 64.3         & 145.6 \\

GLCom  & 15  & --
  & 11,250 & 26.7 & \textbf{100} & 750.0 & 25,000  & \textbf{73.3} & \textbf{100} & 2273.6
  & 81      & 29.3          & 30.7         & \textbf{18.4}
  & 444     & 16.5          & \textbf{100} & 179.4
  & 1,878   & \textbf{73.3} & \textbf{100} & 170.9 \\

GLGrps  & 17  & --
  & 14,450 & 76.5 & 94.1 & 850.0 & 30,000  & 70.6         & \textbf{100}          & 2500.0
  & 86      & 45.9         & 51.8          & \textbf{11.0}
  & 429     & 50.1         & \textbf{100}  & 50.5
  & 1,593   & \textbf{80.0} & \textbf{100}         & 117.1 \\

GLIss  & 27  & --
  & 36,450 & 77.8 & 96.3 & 1350.0 & 20,000  & 94.1         & \textbf{100}          & 787.9
  & 204     & 45.2         & 51.1          & \textbf{16.7}
  & 816     & 37.8         & \textbf{100}  & 80.0
  & 3,617   & \textbf{96.3} & \textbf{100} & 139.1 \\

GLBr  & 9   & --
  & 4,050 & 66.7 & 66.7 & 450.0 & 10,000  & \textbf{100} & \textbf{100} & 1111.1
  & 101     & 71.1         & \textbf{100} & \textbf{15.8}
  & 196     & 53.3         & \textbf{100} & 40.8
  & 803     & \textbf{100} & \textbf{100} & 89.2 \\

GLPro & 31 & --
  & 14,450 & 70.6 & \textbf{100} & 466.1 & 30,000  & \textbf{91.0} & \textbf{100}         & 1063.8
  & 215     & 50.3          & 62.6         & \textbf{13.8}
  & 815     & 64.7          & \textbf{100} & 40.7
  & 4,030   & 90.3          & \textbf{100} & 143.8 \\

GLRepo    & 10  & --
  & 3,200 & 50.0 & 87.5 & 320.0 & 15,000  & 68.0         & \textbf{100} & 2205.9
  & 92      & 66.0         & 93.3         & \textbf{13.9}
  & 230     & 37.2         & \textbf{100} & 61.8
  & 1,056   & \textbf{80.0} & \textbf{100} & 132.0 \\

FDIC           & 6   & --
  & 1,800 & \textbf{100} & \textbf{100} & 300.0 & 5,000   & \textbf{100} & \textbf{100} & 833.3
  & 111     & 90.0         & 93.3         & \textbf{20.6}
  & 152     & \textbf{100} & \textbf{100} & 25.3
  & 321     & \textbf{100} & \textbf{100} & 53.5 \\

PetStore       & 19  & --
  & 2,450 & 85.7 & 13.3 & 128.9 & 100,000 & \textbf{87.4} & \textbf{73.8} & 6024.1
  & 208     & 84.2          & 64.9          & \textbf{13.0}
  & 185     & 66.3          & 68.4          & 14.7
  & 2,391   & 85.2          & 29.8          & 147.8 \\

OhSome         & 134 & --
  & 5,539 & 67.9 & 18.7 & 41.3 & 2,518   & 37.2         & 22.4         & 50.6
  & 2,245   & 95.4 & \textbf{49.5} & \textbf{17.5}
  & 2,355   & 79.6         & 27.0         & 22.1
  & 4,141      & \textbf{98.1} & 29.3 & 31.5 \\

\midrule
\textbf{Average} & 413 & -- 
  & 8,250.6 & 60.4 & 59.4 & 380.8 & 912,518 & 70.4         & 68.2         & 5323.4
  & 4,824   & 79.0         & 69.2         & \textbf{14.6}
  & 7,219   & 68.1         & 65.5         & 36.5
  & 41,551  & \textbf{92.3} & \textbf{67.8} & 129.8 \\
\bottomrule
\end{tabular}
\end{table*}

%% file: sections/rq1_effectiveness.tex
\subsection{Effectiveness and Efficiency (RQ1)}



\vspace{1pt}
\textit{\bf \em Operation Coverage.}
As seen in Table~\ref{tab:operation_coverages}, \tool\ achieves the highest average operation coverage at \textbf{92.3\%}, significantly outperforming AutoRestTest (70.4\%), RESTifAI (79.0\%), KAT (68.1\%), and {\color{custom-red}DeepREST} (60.4\%). At the service level, \tool\ reaches near-full coverage on \textit{Genome Nexus}, \textit{Languagetool}, \textit{Restcountries}, \textit{FDIC}, and multiple GitLab services such as \textit{Branch} and \textit{Issues}. In more complex, interdependent APIs such as \textit{GitLab Groups} and \textit{GitLab Issues}, \tool\ substantially outperforms all competing tools, demonstrating its ability to handle diverse and deeply nested operation dependencies.

A closer look reveals that \tool\ achieves high operation coverage through its ability to identify flexible execution paths via the ODG and the Executable Workflow Graph. Unlike approaches that rely on rigid, linearly ordered dependency chains, \tool\ systematically discovers alternative paths to reach target operations. For example, RESTifAI relies on strict dependency chains such as \code{POST /projects} $\rightarrow$ \code{POST /projects/\{id\}/repository/commits} $\rightarrow$ \code{GET /projects/\{id\}/repository/commits/\{sha\}}, where a failure at any step renders the target endpoint unreachable. AutoRestTest similarly struggles with such stateful scenarios, often incurring high test case counts or failing to construct valid parameter combinations for deeply nested endpoints. In contrast, \tool\ discovers alternative paths, such as GET-based sequences that bypass resource creation prerequisites, enabling robust and accessible coverage for several operations.

Among the baselines, RESTifAI achieves a competitive average operation coverage of 79.0\% but has unstable performance, with notably low coverage on several GitLab services (e.g., 29.3\% for GitLab Commit and 45.2\% for GitLab Issues). This instability stems from its heavy reliance on LLM-generated sequences without any verification, making it sensitive to hallucinated dependencies. AutoRestTest, while more consistent, generally achieves lower coverage than \tool, particularly on large-scale or complex APIs. KAT, despite occasionally performing well on individual services, records the lowest average operation coverage overall (68.1\%), struggling most on APIs with complex multi-operation dependencies. {\color{custom-blue} Finally, the non-LLM baseline DeepREST lag behind other approaches.}


\textit{Status Code Coverage.}
For successful responses, \tool\ leads with an average 2xx coverage of 92.3\%, reflecting effective exploration of valid API behaviors and happy-path scenarios. It also achieves competitive 4xx coverage at 67.8\%, demonstrating its ability to trigger client-side error conditions essential for robustness testing. The combination of high 2xx and 4xx coverage indicates that \tool exercises both the nominal and error-handling behavior of each API, rather than focusing exclusively on one code.

\vspace{1pt}
{\bf \em Source Code Coverage.} 
A key observation from Table~\ref{tab:code_coverage_comparison} is that higher operation coverage does not always translate to higher code coverage. AutoRestTest slightly surpasses \tool\ in operation coverage on certain services, yet yields consistently lower branch, line, and method coverage. This discrepancy stems from its fuzzing strategy, which generates large numbers of semi-random inputs that predominantly trigger 4xx responses without penetrating deeper execution paths.
\tool achieves the highest average method coverage at 58.6\%, outperforming {\color{custom-red}DeepREST (42.7\%)}, AutoRestTest (56.3\%), KAT (53.2\%), and RESTifAI (52.2\%). It also leads in average line coverage at 58.6\%, compared to 55.1\%, 51.1\%, and 52.2\% for the other tools respectively. For branch coverage, it reaches 47.0\% on average, exceeding {\color{custom-red}DeepREST (33.9\%)}, RESTifAI (41.1\%) and KAT (41.3\%), and AutoRestTest~(46.6\%).

These gains are largely attributable to two complementary mechanisms. First, \tool\ employs a systematic, algorithm-driven data generation strategy (Algorithm~\ref{alg:apipilot_overview}, line 11) that explores both required and optional parameter combinations, including edge cases and incomplete inputs. For instance, \code{GET /projects} in the GitLab Commit API supports optional parameters such as \code{id\_after} for pagination, i.e. cases that LLM-based approaches like RESTifAI and KAT frequently overlook, limiting their input diversity and code coverage. Second, \tool's response-driven refinement continuously refines input constraints based on observed runtime failures. For example, although \code{min\_access\_level} in \code{GET /projects} accepts values in \{0, 10, 20, 30, 40, 50\}, values such as 0 may still produce invalid responses; \tool\ detects and suppresses such patterns over successive iterations. Similarly, erroneous LLM-inferred dependencies such as misusing \code{id} from \code{GET /projects} as \code{namespace\_id} in \code{POST /projects}, are identified and corrected, preventing repeated 404 failures.

\input{tables/code_coverage}

\vspace{2pt}
{\bf \em Test Generation Efficiency}. Test generation efficiency is measured as the number of test cases required per successfully covered operation. As shown in Table~\ref{tab:operation_coverages}, \tool\ requires 129.8 test cases per covered operation on average, higher than RESTifAI (14.6) and KAT (36.5), but significantly lower than {\color{custom-red}DeepREST(380.8)}, AutoRestTest (5323.4). RESTifAI and KAT achieve lower Eff values primarily due to conservative exploration strategies that rely on LLM-guided generation or localized mutations, producing inputs close to previously observed patterns. While this reduces test suite size, it also limits coverage of complex or less obvious behaviors, as reflected in KAT's near-zero 5xx error count and its operation coverage 23.3 percentage points below \tool, despite comparable token cost per test case (160.8 vs.\ 154.7). RESTifAI similarly generates fewer test cases but consumes approximately 19,602.9 tokens per test case, roughly 128 times more than \tool\, making it the least token-efficient approach by a wide margin. In both cases, {\em the apparent efficiency of these tools comes at the cost of substantially reduced coverage}.

%% file: tables/code_coverage.tex

\begin{table}[t]
\centering
\footnotesize
\setlength{\tabcolsep}{1.2pt}
\renewcommand{\arraystretch}{1}

\caption{Branch(BC)/line(LC)/method(MC) Coverage (RQ1).}
\label{tab:code_coverage_comparison}
\vspace{-3pt}
\begin{tabular}{l|ccc|ccc|ccc|ccc|ccc}
\toprule
\textbf{Ser.} 
& \multicolumn{3}{c}{\textbf{{\color{custom-red}DeepREST}}}
& \multicolumn{3}{c}{\textbf{AutoRestTest}} 
& \multicolumn{3}{c}{\textbf{RESTifAI}} 
& \multicolumn{3}{c}{\textbf{KAT}} 
& \multicolumn{3}{c}{\textbf{APIPilot}} \\
 & BC & LC & MC 
 & BC & LC & MC 
 & BC & LC & MC 
 & BC & LC & MC 
 & BC & LC & MC \\
\midrule
GN 
& 21.1 & 32.4 & 27.6
& 44.5 & 56.9 & 51.2
& 34.5 & 43.8 & 34.1
& 34.9 & 42.6 & 34.7 
& \textbf{49.4} & \textbf{62.5} & \textbf{59.7} \\

LT  
& 0.1 & 2.9 & 4.3
& 4.5 & 21.4 & 15.1 
& 1.7 & 22.6 & 13.9 
& 4.2 & 20.0 & 16.2 
& \textbf{5.4} & \textbf{30.3} & \textbf{16.9} \\

RC 
& 84.5 & 76.4 & 84.7
& \textbf{91.7} & \textbf{78.0} & \textbf{84.7} 
& 81.4 & 74.2 & 83.5
& 81.4 & 74.2 & 83.5 
& 86.8 & 77.3 & 84.6 \\

JA 
& 29.3 & 42.5 & 45.0
& 54.5 & 74.8 & 74.6 
& 47.5 & 72.2 & 71.1
& 46.4 & 70.5 & 73.4 
& \textbf{55.2} & \textbf{77.3} & \textbf{75.7} \\

PR
& 34.4 & 40.3 & 51.9
& 38.0 & 44.2 & 55.7
& 40.5 & 48.4 & \textbf{58.2}
& 39.8 & 48.1 & 58.0 
& 38.0 & \textbf{45.5} & 56.0 \\

\midrule
\textbf{Avg.}
& 33.9 & 38.9 & 42.7
& 46.6 & 55.1 & 56.3
& 41.1 & 52.2 & 52.2
& 41.3 & 51.1 & 53.2
& \textbf{47.0} & \textbf{58.6} & \textbf{58.6} \\
\bottomrule
\end{tabular}
\end{table}

%% file: sections/rq2_workflow_exploration.tex
\subsection{Workflow Exploration (RQ2)}

\input{tables/workflow_analyst}

Table~\ref{tab:workflow_analysis} reports the number of successfully executed and total generated workflow sequences for each tool, along with their execution success rates. \tool\ generates substantially more execution sequences ({\bf 875} in total) than RESTifAI (413), reflecting its ability to construct multi-step workflows that capture dependencies across operations rather than mapping each test to a single endpoint. Despite this broader exploration, \tool\ maintains a high execution success rate of {\bf 88.1\%}, outperforming RESTifAI ({\bf 83.1\%}), and KAT ({\bf 86.6\%}). DeepREST is excluded from this comparison because it tests operations independently rather than constructing multi-step dependency workflows. This indicates that the additional sequences generated by \tool\ are not speculative, but they are semantically validated and consistent with API constraints.

The advantage is most pronounced on complex, state-dependent services. In GitLab Projects, \tool\ successfully executes 112 workflow sequences compared to only 17 by RESTifAI, and in GitLab Issues and jhipster-app, \tool's test suites are {\bf 3X-4X larger} while maintaining comparable or higher success rates. RESTifAI, by contrast, constructs its test suite in direct correspondence with the number of API operations, limiting it to validating individual endpoints or simple transitions rather than exploring multi-step dependency~chains.

As an example, in the GitLab Commit service, where RESTifAI achieves only 29.3\% operation coverage, consider testing \code{GET /projects/\{id\}/repository/commits/\{sha\} /discussions}. RESTifAI generates a single path:
\code{POST /projects}~$\rightarrow$ \code{POST} \code{/projects/{id}/repository/commits} $\rightarrow$ \code{GET} \code{/projects/{id}/repository/commits/{sha}/discussions}.
This sequence is infeasible in practice: no endpoint supports creating discussions for a commit in this manner. Thus, the path fails and the target operation cannot be tested. \tool\ infers a valid alternative: \code{GET /projects} $\rightarrow$ \code{GET /projects/\{id\}/repository/commits} $\rightarrow$ \code{GET /projects/\{id\}/repo-\allowbreak sitory/commits/\{sha\}/discussions}. This GET-based sequence successfully retrieves existing discussions, allowing the target endpoint to be exercised. More broadly, since RESTifAI generates only a single execution path per operation, any path failure renders that operation untestable. \tool\ mitigates this by retaining the top-$k$ valid workflows per operation, ensuring that intermediate failures do not block coverage of downstream operations.

%% file: tables/workflow_analyst.tex
\begin{table}[t]
\centering
\small
\setlength{\tabcolsep}{1.8pt}
\caption{Workflow Execution (RQ2). {\em Total/Succ/Rate: the total and successful numbers, and success rate.}}
\label{tab:workflow_analysis}
\vspace{-6pt}
\begin{tabular}{l|ccc|ccc|ccc}
\toprule
\textbf{Service}  & \multicolumn{3}{c|}{\textbf{RESTifAI}} & \multicolumn{3}{c|}{\textbf{KAT}} & \multicolumn{3}{c}{\textbf{\tool}} \\
& Succ. & Total & Rate & Succ. & Total & Rate & Succ. & Total & Rate  \\
\midrule
GN   & 22  & 23  & 95.7 & 28  & 31  & 90.3 & 52  & 71  & 73.2 \\
LT  & 2   & 2   & 100  & 2   & 2   & 100  & 2   & 2   & 100  \\
RC   & 20  & 22  & 90.9 & 22  & 22  & 100  & 22  & 22  & 100  \\
JA   & 32  & 38  & 84.2 & 45  & 56  & 80.4 & 103 & 119 & 86.6 \\
PR  & 31  & 35  & 88.6 & 34  & 49  & 69.4 & 43  & 54  & 79.6 \\
\midrule
Canada.H.    & 4   & 4   & 100  & 4   & 4   & 100  & 12  & 12  & 100  \\
Bills API     & 18  & 21  & 85.7 & 24  & 30  & 80.0 & 43  & 48  & 89.6 \\
GLCom  & 4   & 15  & 26.7 & 26  & 28  & 92.9 & 30  & 41  & 73.2 \\
GLGrps  & 8   & 17  & 47.1 & 30  & 32  & 93.8 & 42  & 47  & 89.4 \\
GLIss.  & 13  & 27  & 48.2 & 48  & 48  & 100  & 90  & 94  & 95.7 \\
GLBr.  & 9   & 9   & 100  & 16  & 16  & 100  & 23  & 23  & 100  \\
GLPro. & 17  & 31  & 54.8 & 48  & 59  & 81.4 & 112 & 124 & 90.3 \\
GLRepo  & 7   & 10  & 70.0 & 10  & 18  & 55.6  & 15  & 24  & 62.5 \\
FDIC    & 6   & 6   & 100  & 6   & 6   & 100  & 6   & 6   & 100  \\
PetStore  & 16  & 19  & 84.2 & 20  & 24  & 83.3 & 43  & 54  & 79.6 \\
OhSome   & 134 & 134 & 100  & 121 & 134 & 90.3 & 133 & 134 & 99.3 \\
\midrule
\textbf{Total}  & 343 & 413 & 83.1 & 484 & 559 & 86.6 & \textbf{771} & \textbf{875} & \textbf{88.1} \\
\bottomrule
\end{tabular}

\end{table}

%% file: sections/rq3_error_detection.tex
\subsection{Error Detection (RQ3)}
\input{tables/error_detection}

Table~\ref{tab:err} reports 5xx errors and undocumented status codes detected by each tool across all 16 services.

{\bf \em 5xx Errors.} {\tool} detects 197 unique 5xx errors, outperforming DeepREST (45), AutoRestTest (178), RESTifAI (151), and KAT (8). This advantage is mainly due~to {\tool}'s ability to {\em expose unhandled faults} in the OhSome service. On the other services, {\tool\ still detects more 5xx errors than DeepREST, RESTifAI, and KAT, but fewer than AutoRestTest. This is expected as {\tool} prioritizes dependency-valid workflows and avoids repeatedly exploring invalid or blacklisted values, whereas AutoRestTest's aggressive mutation strategy generates a large volume of inputs that can reveal more raw 5xx errors. However, on OhSome, AutoRestTest requires substantially more time to fuzz effective inputs and detects fewer 5xx errors before timeout. Overall, {\tool}'s strength lies in achieving higher operation and workflow coverage through deeper dependency-valid sequences, rather than maximizing raw 5xx exploration. We plan to improve its 5xx detection by adding more targeted exploration heuristics.

{\bf \em Undocumented Status Codes.} \tool\ detects 271 undocumented responses, fewer than AutoRestTest (472) and slightly below KAT (287), while remaining higher than RESTifAI (198). The reduction relative to AutoRestTest is consistent with \tool's ability to suppress previously failing inputs via the Contextual Resource Pool and refine parameter constraints through its feedback mechanism, reducing the generation of requests that trigger specification-undefined behavior. RESTifAI's lower count reflects its more conservative operation-level generation, which explores a narrower range of input combinations.

The pattern is consistent across complex APIs. On GitLab services and petclinic-rest, \tool\ maintains moderate 5xx detection while keeping undocumented responses low, suggesting that its workflow-driven exploration reaches meaningful fault-triggering paths without excessive specification deviation. Overall, \tool\ achieves a better balance between fault detection and specification adherence than all three baselines.

%% file: tables/error_detection.tex
\begin{table}[t]
\centering
\footnotesize
\setlength{\tabcolsep}{1.5pt}
\caption{{\color{custom-blue} Error detection (5xx, Undocumented) (RQ3).}}
\label{tab:err}
\vspace{-3pt}
\begin{tabular}{l|cc|cc|cc|cc|cc}
\toprule
\multirow{2}{*}{\textbf{Service}} & \multicolumn{2}{c|}{\textbf{DeepREST}} & \multicolumn{2}{c|}{\textbf{AutoRestTest}} & \multicolumn{2}{c|}{\textbf{RESTifAI}} & \multicolumn{2}{c|}{\textbf{KAT}} & \multicolumn{2}{c}{\textbf{APIPilot}} \\
\cline{2-11}
 & 5xx & Undoc & 5xx & Undoc & 5xx & Undoc & 5xx & Undoc & 5xx & Undoc \\
\midrule
GN   & 0  & 15  & 0   & 34    & 0   & 32   & 0   & 19   & 0   & 16   \\
LT   & 0  & 0   & 1   & 2     & 0   & 2    & 0   & 1    & 0   & 2    \\
RC   & 1  & 1   & 1   & 61    & 0   & 14   & 0   & 24   & 0   & 19   \\
JA   & 3  & 45  & 13  & 178   & 10  & 52   & 3   & 61   & 25  & 58   \\
PR   & 32 & 21  & 35  & 23    & 29  & 19   & 5   & 14   & 29  & 4    \\
\midrule
Canada.   & 0 & 1   & 0   & 8     & 0   & 2    & 0   & 2    & 0   & 2    \\
Bills API & 0 & 0   & 1   & 32    & 1   & 0    & 0   & 2    & 0   & 3    \\
GLCom.    & 1 & 1   & 15  & 3     & 0   & 6    & 0   & 17   & 1   & 38   \\
GLGrps.   & 1 & 1   & 17  & 2     & 2   & 10   & 0   & 24   & 6   & 18   \\
GLIss.    & 5 & 10  & 27  & 5     & 2   & 20   & 0   & 44   & 4   & 27   \\
GLBr.     & 0 & 16  & 9   & 28    & 0   & 12   & 0   & 12   & 1   & 24   \\
GLPro.    & 0 & 5   & 31  & 5     & 2   & 19   & 0   & 48   & 7   & 35   \\
GLRepo    & 0 & 10  & 10  & 24    & 0   & 10   & 0   & 11   & 1   & 23   \\
FDIC      & 0 & 6   & 6   & 12    & 3   & 0    & 0   & 0    & 1   & 1    \\
PetStore  & 2 & 2   & 11  & 39    & 6   & 0    & 0   & 8    & 12  & 1    \\
OhSome    & 0 & 0   & 1   & 16    & 96  & 0    & 0   & 0    & 110 & 0    \\
\midrule
\textbf{Total} & 45 & 134 & 178 & \textbf{472} & 151 & 198 & 8 & 287 & \textbf{197} & 271 \\
\bottomrule
\end{tabular}
\vspace{5pt}

\end{table}

%% file: sections/rq4_cost_efficiency.tex
\subsection{Token and Cost Efficiency (RQ4)}

Table~\ref{tab:cost-comparison} reports input tokens, output tokens, and tokens per test case ($T_t$) for each tool across all 16 services. {\color{custom-blue}DeepREST is a non-LLM approach, thus, does not have LLM costs.}

\textit{Token Consumption.} AutoRestTest achieves the lowest token consumption per test case (3), owing to its minimal reliance on LLMs as it generates test cases at scale through mutation and evolutionary search rather than LLM prompting. At the opposite end,~RESTifAI consumes 19,603 tokens per test case, reflecting its heavy dependence on LLM-based generation for every request sequence. \tool\ requires 155 tokens per test case, significantly below RESTifAI and comparable to KAT (161), while generating substantially more tests than both, 41,551 versus 7,219 for KAT and 4,824 for RESTifAI.

\vspace{1pt}
\textit{Monetary Cost.} KAT incurs the lowest total cost (\$0.54), followed by AutoRestTest (\$1.11) and \tool\ (\$3.12), while RESTifAI is much more expensive (\$38.05) (\$0.4 per 1M input tokens, \$0.1 if cached; \$1.6 per 1M output tokens). \tool's higher absolute cost relative to KAT and AutoRestTest is directly attributable to its broader test suite. It generates roughly 6X more test cases than KAT as it utilizes LLM more.


\vspace{1pt}
\textit{Cost-Effectiveness.} Despite its higher absolute cost, \tool\ achieves the best coverage outcomes across all dimensions. It attains the highest average operation coverage (92.3\%),~the best code coverage (47.0\% BC, 58.6\% LC, and 58.6\% MC), and the highest execution success rate (88.1\%). KAT and~RESTifAI, while cheaper per test case, achieve 23.3 and 13.3 percentage points lower operation coverage respectively, meaning their apparent token efficiency comes at the cost of much reduced effectiveness. AutoRestTest, despite low per-test-case token cost, requires over 40X more test cases per percentage point of operation coverage than \tool\ ($\mathit{Eff}$=5323.4 vs. 129.8).

\input{tables/cost_compare}

%% file: tables/cost_compare.tex
\begin{table}[t]
\centering
\footnotesize
\setlength{\tabcolsep}{1pt} 
\caption{Cost–Efficiency Based on Token Usage (RQ4)}
\label{tab:cost-comparison}
\vspace{-6pt}
\begin{tabular}{l|rrr|rrr|rrr|rrr}
\toprule
\multirow{2}{*}{\textbf{Service}} & \multicolumn{3}{c}{\textbf{AutoRestTest}} & \multicolumn{3}{c}{\textbf{RestifAI}} & \multicolumn{3}{c}{\textbf{KAT}} & \multicolumn{3}{c}{\textbf{\tool}} \\
\cmidrule(r){2-4} \cmidrule(lr){5-7} \cmidrule(lr){8-10} \cmidrule(l){11-13}
 & $T_{in}$ & $T_{out}$ & $T_{t}$ & $T_{in}$ & $T_{out}$ & $T_{t}$ & $T_{in}$ & $T_{out}$ & $T_{t}$ & $T_{in}$ & $T_{out}$ & $T_{t}$ \\
\midrule
Genome. & 59 & 9 & 17 & 6055 & 36 & 28,864 & 57 & 54 & 387 & 414 & 230 & 271 \\

Lang.  & 6 & 2 & 0 & 175 & 4 & 5409 & 4 & 4 & 877 & 17 & 13 & 139 \\

Rest. & 44 & 5 & 0 & 1858 & 27 & 8,934 & 4 & 27 & 226 & 92 & 88 & 48 \\

jhipster. & 81 & 20 & 1 & 1,729 & 59 & 4,718 & 85 & 70 & 292 & 361 & 341 & 127 \\

petclinic. & 150 & 13 & 1 & 4,048 & 52 & 10,226 & 65 & 30 & 245 & 412 & 372 & 114 \\

Canada. & 16 & 1 & 1 & 1,773 & 7 & 38,693 & 9 & 3 & 223 & 17 & 12 & 84 \\

Bills & 51 & 8 & 7 & 3,851 & 39 & 19,453 & 28 & 10 & 192 & 122 & 76 & 75 \\

GLCom. & 63 & 19 & 3 & 1,980 & 94 & 25,607 & 32 & 11 & 97 & 117 & 70 & 100 \\

GLGrps & 64 & 16 & 3 & 1,393 & 36 & 16,616 & 49 & 17 & 153 & 126 & 101 & 143 \\

GLIss. & 101 & 32 & 7 & 5,912 & 118 & 29,560 & 93 & 30 & 151 & 330 & 237 & 157 \\

GLBr. & 36 & 11 & 5 & 1,545 & 47 & 15,760 & 22 & 8 & 152 & 85 & 64 & 185 \\

GLPro. & 106 & 34 & 5 & 5,747 & 183 & 27,579 & 94 & 31 & 154 & 391 & 269 & 164 \\

GLRepo & 39 & 13 & 3 & 1,633 & 64 & 18,445 & 29 & 10 & 172 & 226 & 52 & 263 \\

FDIC & 24 & 9 & 6 & 5,007 & 16 & 45,248 & 23 & 6 & 194 & 60 & 26 & 268 \\

PetStore & 42 & 10 & 1 & 607 & 31 & 3,066 & 13 & 5 & 95 & 199 & 195 & 165 \\

OhSome & 1,133 & 314 & 575 & 50,148 & 291 & 22,468 & 178 & 61 & 102 & 70 & 62 & 317 \\
\midrule
\textbf{Total} & 2,013 & 514 & \textbf{3} & 93,461 & 1,104 & 19,603 & \textbf{785} & \textbf{376} & 161 & 3,664 & 2,764 & 155 \\
\bottomrule
\end{tabular}
    $T_{in}$: Input tokens (thousands), $T_{out}$: Output tokens (thousands), $T_{t}$: Tokens per test case.
\end{table}

%% file: sections/threats_to_validity.tex
\section{Threats to Validity}
\label{sec:threats_to_validity}





\textbf{Internal Validity.}
Results may be influenced by implementation differences across \tool\ and the baselines. All LLM-based tools use the same model with temperature 0, but differences in prompts may affect outcomes. Dependency validation and refinement also introduce additional heuristics. Multiple runs and averaging reduce the risk, but residual nondeterminism in LLM outputs and API responses may~remain.



\textbf{External Validity.}
Our benchmark includes 16 real-world APIs from diverse domains, but may not represent all API styles, especially highly stateful, event-driven, or non-RESTful services. Since \tool\ and the baselines rely on OpenAPI specifications, the results may not generalize to poorly documented or specification-inconsistent APIs.

\textbf{Conclusion Validity.}
Baseline selection may affect the observed gains. We compare our tool with recent LLM-based and non-LLM REST API testing approaches closest to ours.

%% file: sections/related.tex
\section{Related Work}
\label{sec:related}

The exist several surveys on RESTful API testing~\cite{kim2022automated,golmohammadi2022testing,ehsan2022restful,martin2022online,sharma2018automated,marculescu2022faults,martin2021black}. Existing approaches can generally be categorized into two directions: \emph{black-box testing}~\cite{viglianisi2020resttestgen,MartinLopez2021Restest,liu2022morest,atlidakis2019restler,laranjeiro2021black,wu2022combinatorial,corradini2022automated} and \emph{white-box testing}~\cite{sahin2021discrete,arcuri2020automated,arcuri2021enhancing,arcuri2020handling,zhang2021adaptive,zhang2021enhancing,zhang2019resource,zhang2021resource}. Black-box approaches typically rely on API specifications such as OpenAPI to derive request structures and dependencies, whereas white-box techniques analyze server-side code to guide test generation.

\textit{Operation dependencies.}
Modeling dependencies among API operations is essential for generating meaningful request sequences. Viglianisi \emph{et al.} introduced the \emph{Operation Dependency Graph} to model data dependencies among operations from OpenAPI specifications~\cite{viglianisi2020resttestgen}. Liu \emph{et al.} proposed the RESTful-service Property Graph (RPG) to represent producer--consumer relationships among operations and dynamically refine dependencies using execution feedback~\cite{liu2022morest}. RESTler~\cite{atlidakis2019restler} infers dependencies 
using a grammar-based approach. Other studies explored structural models such as linkage model trees~\cite{stallenberg2021improving} and API forests~\cite{lin2022forest}. We did not compare
with RESTler and EvoMaster as DeepREST is a more recent tool.

Several works investigate dependencies among parameters within API operations. Martin-Lopez \emph{et al.} proposed the \emph{Inter-parameter Dependency Language} (IDL) to specify parameter constraints explicitly~\cite{martin2020automated}, which is used by RESTest to support constraint-based testing~\cite{MartinLopez2021Restest}. Wu \emph{et al.} extracted inter-parameter constraints from API documentation using pattern-based NLP~\cite{wu2022combinatorial}. While formalizing parameter dependencies, these approaches often require manual annotations or predefined patterns. Recent work explored learning-based approaches, including reinforcement learning~\cite{kim2023adaptive} and deep learning models~\cite{mirabella2021deep}. In contrast, we leverage LLMs to automatically infer such dependencies directly from OpenAPI specifications.

\textit{NLP in API testing.} Classic NLP methods extract testing knowledge from documentation. NLP2REST~\cite{kim2023enhancing} augments OpenAPI specifications with rules inferred from natural language descriptions. Wanwarang~\emph{et al.}~\cite{wanwarang2020testing} and ARTE~\cite{alonso2022arte} generate realistic inputs by querying knowledge bases using semantic concepts extracted from parameter labels and API structure. RESTInfer~\cite{liu2022restinfer} infers parameter constraints from API descriptions through a two-phase analysis.

\textit{LLM-based RESTful API Testing.} Recent work leverages LLMs to exploit semantic information in OpenAPI specifications. A first line enriches or infers specifications: \textsc{RESTSpecIT}~\cite{decrop2025restnowautomatedrest} infers OpenAPI specifications with minimal user input, while \textsc{RESTGPT}~\cite{kim2024leveraging} augments existing ones with machine-interpretable rules and example values. A second line uses LLMs for test generation: \textsc{KAT}~\cite{le2024kat} constructs an ODG and generates test scripts via GPT prompting; \textsc{AutoRestTest}~\cite{kim2025autoresttest} combines LLMs with a semantic dependency graph and multi-agent reinforcement learning; \textsc{LlamaRestTest}~\cite{kim2025llamaresttest} fine-tunes small language models using server responses as feedback; and \textsc{RESTifAI}~\cite{kogler2025restifai} generates reusable CI/CD-ready tests emphasising happy-path workflows. Our work differs in that it treats LLM reasoning as a candidate inference step, verifies inferred dependencies through concrete executions, and generates test sequences deterministically over the verified dependency graph.


%% file: references-kat.bib
@String{Computing = "Computing" }

@String{Springer = "Springer-Verlag" }

@article{sahin2021discrete,
  title={{A discrete dynamic artificial bee colony with hyper-scout for RESTful web service API test suite generation}},
  author={Sahin, Omur and Akay, Bahriye},
  journal={Applied Soft Computing},
  volume={104},
  pages={107246},
  year={2021},
  publisher={Elsevier}
}

@inproceedings{martin2022online,
  title={{Online testing of RESTful APIs: Promises and challenges}},
  author={Martin-Lopez, Alberto and Segura, Sergio and Ruiz-Cort{\'e}s, Antonio},
  booktitle={Proceedings of the 30th ACM Joint European Software Engineering Conference and Symposium on the Foundations of Software Engineering},
  pages={408--420},
  year={2022}
}

@article{ehsan2022restful,
  title={{RESTful API testing methodologies: Rationale, challenges, and solution directions}},
  author={Ehsan, Adeel and Abuhaliqa, Mohammed Ahmad ME and Catal, Cagatay and Mishra, Deepti},
  journal={Applied Sciences},
  volume={12},
  number={9},
  pages={4369},
  year={2022},
  publisher={MDPI}
}

@article{arcuri2020automated,
  title={{Automated black-and white-box testing of restful APIs with evomaster}},
  author={Arcuri, Andrea},
  journal={IEEE Software},
  volume={38},
  number={3},
  pages={72--78},
  year={2020},
  publisher={IEEE}
}

@article{arcuri2021enhancing,
  title={{Enhancing search-based testing with testability transformations for existing APIs}},
  author={Arcuri, Andrea and Galeotti, Juan P},
  journal={ACM Transactions on Software Engineering and Methodology (TOSEM)},
  volume={31},
  number={1},
  pages={1--34},
  year={2021},
  publisher={ACM New York, NY}
}

@inproceedings{wanwarang2020testing,
  title={Testing apps with real-world inputs},
  author={Wanwarang, Tanapuch and Borges Jr, Nataniel P and Bettscheider, Leon and Zeller, Andreas},
  booktitle={Proceedings of the IEEE/ACM 1st International Conference on Automation of Software Test},
  pages={1--10},
  year={2020}
}

@article{lin2022forest,
  title={{foREST: A Tree-based Approach for Fuzzing RESTful APIs}},
  author={Lin, Jiaxian and Li, Tianyu and Chen, Yang and Wei, Guangsheng and Lin, Jiadong and Zhang, Sen and Xu, Hui},
  journal={arXiv preprint arXiv:2203.02906},
  year={2022}
}

@article{kim2023adaptive,
  title={{Adaptive REST API Testing with Reinforcement Learning}},
  author={Kim, Myeongsoo and Sinha, Saurabh and Orso, Alessandro},
  journal={arXiv preprint arXiv:2309.04583},
  year={2023}
}

@article{arcuri2020handling,
  title={Handling SQL databases in automated system test generation},
  author={Arcuri, Andrea and Galeotti, Juan P},
  journal={ACM Transactions on Software Engineering and Methodology (TOSEM)},
  volume={29},
  number={4},
  pages={1--31},
  year={2020},
  publisher={ACM New York, NY, USA}
}

@inproceedings{sharma2018automated,
  title={{Automated API testing}},
  author={Sharma, Abhinav and Revathi, M and others},
  booktitle={2018 3rd International Conference on Inventive Computation Technologies (ICICT)},
  pages={788--791},
  year={2018},
  organization={IEEE}
}

@inproceedings{kim2022automated,
  title={Automated test generation for rest apis: No time to rest yet},
  author={Kim, Myeongsoo and Xin, Qi and Sinha, Saurabh and Orso, Alessandro},
  booktitle={Proceedings of the 31st ACM SIGSOFT International Symposium on Software Testing and Analysis},
  pages={289--301},
  year={2022}
}

@article{zhang2021adaptive,
  title={{Adaptive hypermutation for search-based system test generation: A study on rest APIs with EvoMaster}},
  author={Zhang, Man and Arcuri, Andrea},
  journal={ACM Transactions on Software Engineering and Methodology (TOSEM)},
  volume={31},
  number={1},
  pages={1--52},
  year={2021},
  publisher={ACM New York, NY}
}

@inproceedings{zhang2021enhancing,
  title={{Enhancing resource-based test case generation for RESTful APIs with SQL handling}},
  author={Zhang, Man and Arcuri, Andrea},
  booktitle={International Symposium on Search Based Software Engineering},
  pages={103--117},
  year={2021},
  organization={Springer}
}

@inproceedings{zhang2019resource,
  title={{Resource-based test case generation for RESTful web services}},
  author={Zhang, Man and Marculescu, Bogdan and Arcuri, Andrea},
  booktitle={Proceedings of the genetic and evolutionary computation conference},
  pages={1426--1434},
  year={2019}
}

@article{zhang2021resource,
  title={Resource and dependency based test case generation for RESTful Web services},
  author={Zhang, Man and Marculescu, Bogdan and Arcuri, Andrea},
  journal={Empirical Software Engineering},
  volume={26},
  number={4},
  pages={76},
  year={2021},
  publisher={Springer}
}

@Inproceedings{viglianisi2020resttestgen,
  title={{RestTestGen: automated black-box testing of restful APIs}},
  author={Viglianisi, Emanuele and Dallago, Michael and Ceccato, Mariano},
  booktitle={2020 IEEE 13th International Conference on Software Testing, Validation and Verification (ICST)},
  pages={142--152},
  year={2020},
  organization={IEEE}
}

@inproceedings{martin2020automated,
  title={{Automated analysis of inter-parameter dependencies in web APIs}},
  author={Martin-Lopez, Alberto},
  booktitle={Proceedings of the ACM/IEEE 42nd International Conference on Software Engineering: Companion Proceedings},
  pages={140--142},
  year={2020}
}

@article{laranjeiro2021black,
  title={A black box tool for robustness testing of REST services},
  author={Laranjeiro, Nuno and Agnelo, Jo{\~a}o and Bernardino, Jorge},
  journal={IEEE Access},
  volume={9},
  pages={24738--24754},
  year={2021},
  publisher={IEEE}
}

@inproceedings{atlidakis2019restler,
  title={{Restler: Stateful rest API fuzzing}},
  author={Atlidakis, Vaggelis and Godefroid, Patrice and Polishchuk, Marina},
  booktitle={2019 IEEE/ACM 41st International Conference on Software Engineering (ICSE)},
  pages={748--758},
  year={2019},
  organization={IEEE}
}

@inproceedings{liu2022morest,
  title={Morest: model-based RESTful API testing with execution feedback},
  author={Liu, Yi and Li, Yuekang and Deng, Gelei and Liu, Yang and Wan, Ruiyuan and Wu, Runchao and Ji, Dandan and Xu, Shiheng and Bao, Minli},
  booktitle={Proceedings of the 44th International Conference on Software Engineering},
  pages={1406--1417},
  year={2022}
}

@inproceedings{liu2022restinfer,
  title={{RESTInfer: automated inferring parameter constraints from natural language RESTful API descriptions}},
  author={Liu, Yi},
  booktitle={Proceedings of the 30th ACM Joint European Software Engineering Conference and Symposium on the Foundations of Software Engineering},
  pages={1816--1818},
  year={2022}
}

@article{alonso2022arte,
  title={{ARTE: Automated Generation of Realistic Test Inputs for Web APIs}},
  author={Alonso, Juan C and Martin-Lopez, Alberto and Segura, Sergio and Garcia, Jose Maria and Ruiz-Cortes, Antonio},
  journal={IEEE Transactions on Software Engineering},
  volume={49},
  number={1},
  pages={348--363},
  year={2022},
  publisher={IEEE}
}

@article{golmohammadi2022testing,
  title={{Testing RESTful APIs: A Survey}},
  author={Golmohammadi, Amid and Zhang, Man and Arcuri, Andrea},
  journal={ACM Transactions on Software Engineering and Methodology},
  year={2022},
  publisher={ACM New York, NY}
}

@Inproceedings{kim2023enhancing,
  title={{Enhancing REST API Testing with NLP Techniques}},
  author={Kim, Myeongsoo and Corradini, Davide and Sinha, Saurabh and Orso, Alessandro and Pasqua, Michele and Tzoref-Brill, Rachel and Ceccato, Mariano},
  booktitle={Proceedings of the 32nd ACM SIGSOFT International Symposium on Software Testing and Analysis},
  pages={1232--1243},
  year={2023}
}

@inproceedings{wu2022combinatorial,
  title={{Combinatorial testing of restful APIs}},
  author={Wu, Huayao and Xu, Lixin and Niu, Xintao and Nie, Changhai},
  booktitle={Proceedings of the 44th International Conference on Software Engineering},
  pages={426--437},
  year={2022}
}

@inproceedings{MartinLopez2021Restest,
title= {{RESTest: Automated Black-Box Testing of RESTful Web APIs}},
author= {Alberto Martin-Lopez and Sergio Segura and Antonio Ruiz-Cort\'{e}s},
booktitle= {Proceedings of the 30th ACM SIGSOFT International Symposium on Software Testing and Analysis},
series= {ISSTA '21},
publisher= {Association for Computing Machinery},
year= {2021}
}

@inproceedings{mirabella2021deep,
  title={{Deep learning-based prediction of test input validity for restful APIs}},
  author={Mirabella, A Giuliano and Martin-Lopez, Alberto and Segura, Sergio and Valencia-Cabrera, Luis and Ruiz-Cort{\'e}s, Antonio},
  booktitle={2021 IEEE/ACM Third International Workshop on Deep Learning for Testing and Testing for Deep Learning (DeepTest)},
  pages={9--16},
  year={2021},
  organization={IEEE}
}

@inproceedings{martin2021black,
  title={{Black-box and white-box test case generation for RESTful APIs: Enemies or allies?}},
  author={Martin-Lopez, Alberto and Arcuri, Andrea and Segura, Sergio and Ruiz-Cort{\'e}s, Antonio},
  booktitle={2021 IEEE 32nd International Symposium on Software Reliability Engineering (ISSRE)},
  pages={231--241},
  year={2021},
  organization={IEEE}
}

@article{marculescu2022faults,
  title={{On the faults found in rest APIs by automated test generation}},
  author={Marculescu, Bogdan and Zhang, Man and Arcuri, Andrea},
  journal={ACM Transactions on Software Engineering and Methodology (TOSEM)},
  volume={31},
  number={3},
  pages={1--43},
  year={2022},
  publisher={ACM New York, NY}
}

@article{corradini2022automated,
  title={{Automated black-box testing of nominal and error scenarios in RESTful APIs}},
  author={Corradini, Davide and Zampieri, Amedeo and Pasqua, Michele and Viglianisi, Emanuele and Dallago, Michael and Ceccato, Mariano},
  journal={Software Testing, Verification and Reliability},
  volume={32},
  number={5},
  pages={e1808},
  year={2022},
  publisher={Wiley Online Library}
}

@inproceedings{stallenberg2021improving,
  title={{Improving test case generation for rest APIs through hierarchical clustering}},
  author={Stallenberg, Dimitri and Olsthoorn, Mitchell and Panichella, Annibale},
  booktitle={2021 36th IEEE/ACM International Conference on Automated Software Engineering (ASE)},
  pages={117--128},
  year={2021},
  organization={IEEE}
}


%% file: references.bib
@phdthesis{fielding2000rest,
  title={Architectural styles and the design of network-based software architectures},
  author={Fielding, Roy Thomas},
  year={2000},
  school={University of California, Irvine}
}

@article{godefroid2012sage,
author = {Godefroid, Patrice and Levin, Michael Y. and Molnar, David},
title = {SAGE: whitebox fuzzing for security testing},
year = {2012},
issue_date = {March 2012},
publisher = {Association for Computing Machinery},
address = {New York, NY, USA},
volume = {55},
number = {3},
issn = {0001-0782},
url = {https://doi.org/10.1145/2093548.2093564},
doi = {10.1145/2093548.2093564},
journal = {Commun. ACM},
month = mar,
pages = {40–44},
numpages = {5}
}

@INPROCEEDINGS{arcuri2018evomaster,
  author={Arcuri, Andrea},
  booktitle={2018 IEEE 11th International Conference on Software Testing, Verification and Validation (ICST)}, 
  title={EvoMaster: Evolutionary Multi-context Automated System Test Generation}, 
  year={2018},
  volume={},
  number={},
  pages={394-397},
  doi={10.1109/ICST.2018.00046}}

@misc{kogler2025restifai,
      title={RESTifAI: LLM-Based Workflow for Reusable REST API Testing}, 
      author={Leon Kogler and Maximilian Ehrhart and Benedikt Dornauer and Eduard Paul Enoiu},
      year={2025},
      eprint={2512.08706},
      archivePrefix={arXiv},
      primaryClass={cs.SE},
      url={https://arxiv.org/abs/2512.08706}, 
}

@inproceedings{kim2025autoresttest,
author = {Kim, Myeongsoo and Stennett, Tyler and Sinha, Saurabh and Orso, Alessandro},
title = {A Multi-Agent Approach for REST API Testing with Semantic Graphs and LLM-Driven Inputs},
year = {2025},
isbn = {9798331505691},
publisher = {IEEE Press},
url = {https://doi.org/10.1109/ICSE55347.2025.00179},
doi = {10.1109/ICSE55347.2025.00179},
booktitle = {Proceedings of the IEEE/ACM 47th International Conference on Software Engineering},
pages = {1409–1421},
numpages = {13},
location = {Ottawa, Ontario, Canada},
series = {ICSE '25}
}

@misc{decrop2025restnowautomatedrest,
      title={You Can REST Now: Automated REST API Documentation and Testing via LLM-Assisted Request Mutations}, 
      author={Alix Decrop and Xavier Devroey and Mike Papadakis and Pierre-Yves Schobbens and Gilles Perrouin},
      year={2025},
      eprint={2402.05102},
      archivePrefix={arXiv},
      primaryClass={cs.SE},
      url={https://arxiv.org/abs/2402.05102}, 
}

@inproceedings{kim2024leveraging,
author = {Kim, Myeongsoo and Stennett, Tyler and Shah, Dhruv and Sinha, Saurabh and Orso, Alessandro},
title = {Leveraging Large Language Models to Improve REST API Testing},
year = {2024},
isbn = {9798400705007},
publisher = {Association for Computing Machinery},
address = {New York, NY, USA},
url = {https://doi.org/10.1145/3639476.3639769},
doi = {10.1145/3639476.3639769},
booktitle = {Proceedings of the 2024 ACM/IEEE 44th International Conference on Software Engineering: New Ideas and Emerging Results},
pages = {37–41},
numpages = {5},
location = {Lisbon, Portugal},
series = {ICSE-NIER'24}
}

@INPROCEEDINGS{le2024kat,
  author={Le, Tri and Tran, Thien and Cao, Duy and Le, Vy and Nguyen, Tien N. and Nguyen, Vu},
  booktitle={2024 IEEE Conference on Software Testing, Verification and Validation (ICST)}, 
  title={KAT: Dependency-Aware Automated API Testing with Large Language Models}, 
  year={2024},
  volume={},
  number={},
  pages={82-92},
  doi={10.1109/ICST60714.2024.00017}}

@article{kim2025llamaresttest,
author = {Kim, Myeongsoo and Sinha, Saurabh and Orso, Alessandro},
title = {LlamaRestTest: Effective REST API Testing with Small Language Models},
year = {2025},
issue_date = {July 2025},
publisher = {Association for Computing Machinery},
address = {New York, NY, USA},
volume = {2},
number = {FSE},
url = {https://doi.org/10.1145/3715737},
doi = {10.1145/3715737},
journal = {Proc. ACM Softw. Eng.},
month = jun,
articleno = {FSE022},
numpages = {24}
}

@misc{jhipster,
  author = {{JHipster Contributors}},
  title = {JHipster: A development platform to quickly generate, develop, and deploy modern web applications and microservice architectures},
  year = {2026},
  publisher = {GitHub},
  journal = {GitHub repository},
  howpublished = {\url{https://github.com/jhipster/generator-jhipster}},
  note = {Accessed: 2026-03-22}
}

@misc{spring-petclinic,
  author = {{Spring Community}},
  title = {Spring PetClinic Sample Application},
  year = {2026},
  publisher = {GitHub},
  journal = {GitHub repository},
  howpublished = {\url{https://github.com/spring-projects/spring-petclinic}},
  note = {Accessed: 2026-03-22}
}

@article{neumann2018analysis,
  title={An analysis of public REST web service APIs},
  author={Neumann, Andy and Laranjeiro, Nuno and Bernardino, Jorge},
  journal={IEEE Transactions on Services Computing},
  volume={14},
  number={4},
  pages={957--970},
  year={2018},
  publisher={IEEE}
}

@article{zhang2023open,
  title={Open problems in fuzzing {RESTful} {APIs}: A comparison of tools},
  author={Zhang, Man and Arcuri, Andrea},
  journal={ACM Transactions on Software Engineering and Methodology (TOSEM)},
  volume={32},
  number={6},
  pages={1--40},
  year={2023},
  publisher={ACM New York, NY, USA}
}

@inproceedings{zhang2025hallucinations,
  title={{LLM Hallucinations in Practical Code Generation: Phenomena, Mechanism, and Mitigation}},
  author={Zhang},
  booktitle={Proceedings of the ACM on Software Engineering (ISSTA)},
  year={2025},
  publisher={ACM},
  note={Identifies API Knowledge Conflicts as a specific hallucination type}
}

@misc{gitlab,
  author = {{GitLab Inc.}},
  title = {GitLab Community Edition},
  howpublished = {\url{https://gitlab.com/gitlab-org/gitlab}}
}

@misc{api_pilot_replication_2026,
  title = {{APIPilot Replication package}},
  year = {2026},
  howpublished = {\url{https://anonymous.4open.science/r/APIPilot}},
  note = {Jun 2026}
}

@misc{jacoco,
  title={JaCoCo},
  url={https://github.com/jacoco/jacoco},
  year={2026}
}

@inproceedings{corradini2024DeepREST,
author = {Corradini, Davide and Montolli, Zeno and Pasqua, Michele and Ceccato, Mariano},
title = {DeepREST: Automated Test Case Generation for REST APIs Exploiting Deep Reinforcement Learning},
year = {2024},
isbn = {9798400712487},
publisher = {Association for Computing Machinery},
address = {New York, NY, USA},
url = {https://doi.org/10.1145/3691620.3695511},
doi = {10.1145/3691620.3695511},
booktitle = {Proceedings of the 39th IEEE/ACM International Conference on Automated Software Engineering},
pages = {1383–1394},
numpages = {12},
location = {Sacramento, CA, USA},
series = {ASE '24}
}
